# Autonomic Management for Multi-agent Systems

Nadir K.Salih [1#]  Tianyi Zang [1@]  G.K.Viju [2*]  Abdelmotalib A.Mohamed [1&]

[1] School of Computer Science and Engineering, Harbin Institute of Technology, China
[2] Department of Computer Science, Karary University, Khartoum, Sudan

**Abstract-**
*Autonomic computing is a computing system that can manage itself by self-configuration, self-healing, self-optimizing and self-protection. Researchers have been emphasizing the strong role that multi agent systems can play progressively towards the design and implementation of complex autonomic systems. The important of autonomic computing is to create computing systems capable of managing themselves to a far greater extent than they do today. With the nature of autonomy, reactivity, sociality and pro-activity, software agents are promising to make autonomic computing system a reality. This paper mixed multi-agent system with autonomic feature that completely hides its complexity from users/services. Mentioned Java Application Development Framework as platform example of this environment, could applied to web services as front end to users. With multi agent support it also provides adaptability, intelligence, collaboration, goal oriented interactions, flexibility, mobility and persistence in software systems.*

## I. Introduction

A new computational framework called Agent Oriented Programming (AOP), which can be viewed as a specialization of object oriented programming. It is relatively a new software paradigm that brings concepts from the theories of artificial intelligence into the mainstream realm of distributed systems. AOP essentially models an application as a collection of components called agents that are characterized by, among other things, autonomy, proactivity and an ability to communicate. Being autonomous they can independently carry out complex, and often long-term, tasks [1]. Intelligent autonomous agent must build and maintain a model of the external environment and of its own components. Atop-level executive component makes decisions based on the models and its current emotional state. A planner component is used to create multiple step scripts or sequences of actions necessary to achieve the high-level goals being pursued by the executive [2]. Agents have the capability to move from one environment to another see fig.1.In the agent design, using FraMaS "advanced behavior" wrappers (like autonomously search according to the agent knowledge of the user or planning strategy to arrive to the target point) [3]. The autonomy of each agent and the messaging interface are useful in most of flexible and extensible systems. Because agents are not directly linked to others, then it is easy to take one out of operation or add a new one while the others are running [4]. Multi-agent development has emerged as a viable approach to meet the autonomic system requirements-autonomy, adaptability, intelligence, goal-oriented interaction, collaboration, and flexibility. Using multiagent approach, real-world problems can be modeled in the form of autonomous, interacting agent components [8].

An autonomic system is an autonomous computing environment that completely hides its complexity. Complexity hiding from users/services means that autonomic computing will provide users with a computing environment that allows them to concentrate on what they want to do without worrying about how it has to be done [12]. The characteristics of Multiagent Systems (MASs) are that (1) each agent has incomplete information or capabilities for solving the problem and, thus, has a limited viewpoint; (2) there is no system global control; (3) data are decentralized; and (4) computation is asynchronous [13].

The paper is organized as follows: Section II reviews Related Work Section III focuses on the Programming Language and Tools. Section IV describes JADE and the Agents Paradigm. Finally, Section V takes The Utility of Agent and WEB Service Integration before concluding the paper.

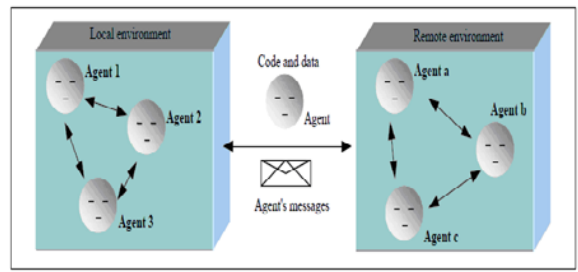

Fig.1 Multi Agent System Model

## II. Related Work

The Unity system components are implemented as autonomic elements— individual agents that control resources and deliver services to humans and to other autonomic elements. Every Unity component is an autonomic element. This includes: computing resources (e.g. databases, storage systems, servers), higher-level elements with some management authority (e.g. workload managers or provisioners), and elements that assist other elements in doing their tasks (e.g. policy repositories, sentinels, brokers, or registries) [9]. Each autonomic element is responsible for its own internal autonomic behavior, namely, managing the resources that it controls, and managing its own internal operations, including self-configuration, self-optimization, self-protection and self-healing. MAACE emphasis on self-organization and self-healing of application services and it is an open and extensible computing environment to allow heterogeneous agent to join it. By the cooperation of agent federation system, agent mediate system and agent monitor system, MAACE lead to automated control and management of a wide range of network centric applications and services [10].The Bean Generator is a tool that supports agent engineers in creating message content ontologies compliant with the JADE support. The tool is a plug-in for Protege, which is a commonly used ontology editor that enables engineers to graphically model ontologies. Furthermore, additional functionality and storage formats can be 'plugged in' to the system. Another advantage of the Protege tool is that other ontologies can be imported. Repositories of existing ontologies ranging from biological domains to market place product and service descriptions can be found at the Protege community page and at the DAML site1. The languages used to represent these ontologies can be XML, RDF, DAML-OIL, XMI, SQL or UML [5]. The jademx JADE add-on, which provides two major capabilities: the ability to interface JADE agents with Java JMX (Java Management Extensions) and the ability to unit test JADE agents using JUnit. Jademx is available for download from the third-party software area of the JADE website. Everyday, useful software systems rarely exist in isolation. Indeed, one of the strengths of JADE is that the full capabilities of the Java environment are available when creating a software agent application. JMX is the Java technology for management and monitoring of software systems; it was originally part of the Java EE enterprise platform (formerly known as J2EE), but as of Java 5 it is available as part of the standard J2SE environment. Furthermore, unit testing is an important technique for the development of robust software and Junit is a standard methodology for the unit testing of applications written in Java. Jademx was developed for an industrial software agent effort requiring management using Java EE and to be unit-testable. A jademx agent can be configured either programmatically or by using XML [1]. The Java Sniffer is a stand-alone Java application, developed by Rockwell Automation, Inc., that can remotely connect to running JADE systems and is intended as an alternative to JADE's built-in sniffer. The tool receives messages from all agents in the system, reasons about the information, and presents it from different points of view (see Fig.2) [6]. We observed Jadex Belief Desire Intention BDI reasoning engine that allows development of rational agents using mentalistic notions in the implementation layer. In other words, it enables the construction of rational agents following the BDI model. In contrast to all other available BDI engines, Jadex fully supports the two-step practical reasoning process (goal deliberation and means–end reasoning) instead of operationalizing only the means–end process. This means that Jadex allows the construction of agents with explicit representation of mental attitudes (beliefs, goals and plans) and that automatically deliberate about their goals and subsequently pursue them by applying appropriate plans. The reasoning engine is clearly separated from its underlying infrastructure, which provides basic platform services such as life-cycle management and communication. Hence, running Jadex over JADE combines the strength of a well-tested agent middleware with the abstract BDI execution model. For the programming of agents, the engine relies on established techniques, such as Java and XML and, to further simplify the development task, Jadex includes a rich suite of run-time tools that are based upon the JADE administration and debugging tools. It also includes a library of ready-to-use generic functionalities provided by several agent modules (capabilities) [7]. OMACS is a metamodel for agent organizations. It defines the required organizational structure that allows multiagent teams to autonomously reconfigure at runtime, thus enabling them to cope with unpredictable situations in a dynamic environment [11]. MAGE, an agent-oriented programming environment, with complete tools to support agent-based requirement analysis, design, development and deployment, is a powerful development environment for autonomous computing [14].

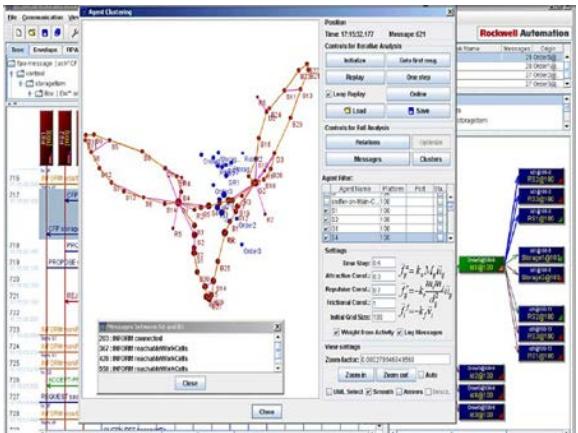

Fig 2 JavaSniffer user interface

### III. Programming Language and Tools

Multi-agent systems programming languages, platforms and development tools are important components that can affect the diffusion and use of agent technologies across different application domains. In fact, the success of multi-agent systems is largely dependent on the availability of appropriate technology (i.e. programming languages, software libraries and development tools) that allows relatively straightforward implementation of the concepts and techniques that form the basis of multi-agent systems. Multi-agent systems can be realized by using any kind of programming language. In particular, object-oriented languages are considered a suitable means because the concept of agent is not too distant from the concept of object. In fact, agents share many properties with objects such as encapsulation, and frequently, inheritance and message passing. However, agents also differ from objects in several key ways; they are autonomous (i.e. they decide for themselves whether or not to perform an action on request from another agent); they are capable of a flexible behavior; and each agent of a system has its own thread of control. An important characteristic that multi-agent systems should provide is the capability to support interoperability among legacy software systems. Therefore, the availability of software tools for their integration with other common technologies can be a key to their success. The Internet is one the most important application domains and the most important communication means that multi-agent systems can use to provide interoperability among legacy software systems; therefore, a lot of current research and development work is oriented towards providing suitable techniques and software tools for the integration of multi-agent systems with Web technologies such as, for example, Web services and Semantic Web technologies.

### IV. JADE and the Agents Paradigm

JADE is a software platform that provides basic middleware-layer functionalities which are independent of the specific application and which simplify the realization of distributed applications that exploit the software agent abstraction. A significant merit of JADE is that it implements this abstraction over a well-known object-oriented language, Java, providing a simple and friendly API. The following simple design choices were influenced by the agent abstraction**.** An Agent is Autonomous and Proactive: An agent cannot provide call-backs or its own object reference to other agents in order to mitigate any chance of other entities coopting control of its services. An agent must have its own thread of execution, using it to control its life cycle and decide autonomously when to perform which actions. The System is Peer-to-Peer each agent is identified by a globally unique name (the Agent Identifier, or AID, as defined by FIPA). It can join and leave a host platform at any time and can discover other agents through both white-page and yellow-page services (provided in JADE by AMS and the DF agents as defined also by the FIPA specifications). An agent can initiate communication with any other agent at any time it wishes and can equally be the object of an incoming communication at any time. On the basis of these design choices, JADE was implemented to provide programmers with the following ready-to-use and easy-to-customize core functionalities:-

• A fully distributed system inhabited by agents, each running as a separate thread, potentially on different remote machines, and capable of transparently communicating with one another, i.e. the platform provides a unique location-independent API that abstracts the underlying communication infrastructure.

• Efficient transport of asynchronous messages via a location-transparent API. The platform selects the best available means of communication and, when possible, avoids marshalling/unmarshalling java objects. When crossing platform boundaries, messages are automatically transformed from JADE's own internal Java representation into proper FIPA-compliant syntaxes, encodings and transport protocols.

• Support for agent mobility. Both agent code and, under certain restrictions, agent state can migrate between processes and machines. Agent migration is made transparent to communicating agents that can continue to interact even during the migration process.

• A set of graphical tools to support programmers when debugging and monitoring. These are particularly important and complex in multi-threaded, multi-process, multi-machine systems such as a typical JADE application.

• Integration with various Web-based technologies including JSP, servlets, applets and Web service

technology. The platform can also be easily configured to cross firewalls and use NAT systems.

• An in-process interface for launching/controlling a platform and its distributed components from an external application.

## V. The Utility of Agent and WEB Service Integration

Integrating Web services and software agents brings about an obvious benefit: connecting application domains by enabling a Web service to invoke an agent service and vice versa. However, this interconnection is more than simply cross-domain discovery and invocation; it will also allow complex compositions of agent services and Web services to be created, managed and administered by controller agents. To the users of Web services, whether human or computational, agents can be a powerful means of indirection by masking the Web service for purposes of, for example, redirection, aggregation, integration and administration. Redirection describes the case where a Web service may no longer be available for some reason, or the owner of the Web service wishes to temporarily redirect invocations to another Web service without removing the original implementation. Aggregation allows several Web services to be composed into logically interconnected clusters, providing patterned abstractions of behavior that can be invoked through a single service interface. Integration describes the means of simply making Web services available to consumers already using, or planning to use, agent platforms for their business applications, and administration covers aspects of automated Web service management where the agent autonomously administers one or more Web services without necessary intervention from a human user.

## VI. Conclusion

Many researchers in the MAS community have recognized the advantages of an agent based approach to building deployable solutions in the number of application domains comprising complex, distributed systems. Autonomic Computing is providing new vistas in reducing the complexity incurred in today's distributed systems. It minimizes human intervention and reduces the administration cost of enterprise IT systems. With multi agent support it also provides adaptability, intelligence, collaboration, goal oriented interactions, flexibility, mobility and persistence in software systems.

In this paper, we mentioned JADE was implemented to provide programmers with the ready-to-use and easy-to-customize core functionalities .An Agent is Autonomous and Proactive. In addition we have recommended an agent-Web service that has the features of both the agent technology as well as the Web services technology and is managed by an autonomic system based on multi-agent support. This can help to develop enterprise IT systems that are optimal, highly available.

References
[1] John Wiley, Sons Ltd, 2007, Developing Multi-Agent Systems with JADE
[2] J. P. Bigus D. A. Schlosnagle, A toolkit for building multiagent autonomic systems, IBM Systems Journal, Vole 41, NO 3, 2002
[3] Henri Avancini, Analía Amandi. A Java Framework for Multi-agent Systems, SADIO Electronic Journal of Informatics and Operations Research, vol. 3, no. 1, pp. 1-12 (2000).
[4] Fatemeh Daneshfar, Hassan Bevrani. Multi-Agent Systems in Control Engineering: A Survey, Hindawi Publishing Corporation Journal of Control Science and Engineering Volume 2009, Article ID 531080, 12 pages.
[5] Alessio Bosca, Dario Bonino. Ontology Exploration through Logical Views in Protégé, $18^{th}$ International Workshop on Database and Expert Systems Applications, IEEE, 2007.
[6] Pavel Vrba, Pavel Tich. Rockwell Automation's Holonic and Multiagent Control Systems Compendium, IEEE Transactions On Systems, MAN, And Cybernetics—PART C: Applications And Reviews, Vole. 41, NO. 1, January 2011
[7] Frank Chiang, Robin Braun, Autonomic Service Configuration for Telec-munication MASS with Extended Role-Based GAIA and JADEx. 2005 IEEE
[8] Gilda Pour, Multi-Agent Autonomic Architectures for Quality Control Systems, San Jose State University San Jose, CA, U.S.A.
[9] Gerald Tesauro, David M. Chess. A Multi-Agent Systems Approach to Autonomic Computing, AAMAS'04, July 19-23, 2004, New York, New York, USA
[10] Jun W, JI Gao, Bei-Shui Liao, Jiu-Jun Chen. Multi - Agent System Based Autonomic Computing Environment, Proceedings of the Third International Conference on Machine Laming and Cybemetics, Shanghai, 26-29 August 2004.
[11] Walamitien H. Oyenan and Scott A. DeLoach. Design and Evaluation of a Multiagent Autonomic Information System. International Conference on Intelligent Agent Technology 2007 IEEE/WIC/ACM.
[12] Huaglory Tianfield. Multi-Agent Autonomic Architecture and Its Application in E- Medicine, Proceedings of the IEEE/WIC International Conference on Intelligent Agent Technology, 2003.
[13] Katia P. Sycara. Multiagent Systems, This Is a Publication of the American Association for Artificial Intelligence, 1998
[14] Zhongzhi Shi, Haijun Zhang, Yong Cheng. MAGE: An Agent-Oriented Software Engineering Environment, Proceedings of the Third IEEE International Conference on Cognitive Informatics, 2004.